\documentclass{pasj00}

\begin{document}
\SetRunningHead{D. Nogami et al.}{A new SU UMa star, QW Ser}

\Received{2002/00/00}
\Accepted{2002/00/00}

\title{A New SU UMa-Type Dwarf Nova, QW~Serpentis (= TmzV46)}

\author{Daisaku \textsc{Nogami}}
\affil{Hida Observatory, Kyoto University,
       Kamitakara, Gifu 506-1314}
\email{nogami@kwasan.kyoto-u.ac.jp}
\author{Makoto \textsc{Uemura}, Ryoko \textsc{Ishioka}, Hidetoshi
\textsc{Iwamatsu}, Taichi \textsc{Kato}}
\affil{Department of Astronomy, Faculty of Science, Kyoto University,
Sakyo-ku, Kyoto 606-8502}
\author{Elena P. \textsc{Pavlenko}}
\affil{Crimean Astrophysical Observatory, Nauchny, 98409 Crimea}
\affil{Isac Newton Institute of Chile, Crimean Branch, Ukraine}
\author{Alex \textsc{Baklanov}}
\affil{Department of Astronomy, Odessa National University,
T. G. Shevchenko park, Odessa 65014, Ukraine}
\author{Rudolf \textsc{Nov\'ak}}
\affil{Nicholas Copernicus Observatory, Kravihora 2, Brno 616 00,
Czech Republic}
\author{Seiichiro \textsc{Kiyota}}
\affil{Variable Star Observers League in Japan (VSOLJ), 1-401-810 Azuma,
Tsukuba, Ibaraki 305-0031}
\author{Kenji \textsc{Tanabe}}
\affil{Department of Biosphere-Geosphere Systems, Faculty of
Informatics, Okayama University of Science, 1-1 Ridaicho,}
\affil{Okayama 700-0005}
\author{Gianluca \textsc{Masi}}
\affil{Physics Department, University of Rome "Tor Vergata" Via
              della Ricerca Scientifica, 1 00133 Rome, Italy}
\author{Lewis M. \textsc{Cook}}
\affil{Center for Backyard Astrophysics (Concord), 1730 Helix Court,
Concord, CA 94518, USA}
\author{Koichi \textsc{Morikawa}}
\affil{VSOLJ, 468-3 Satoyamada, Yakage-cho, Oda-gun, Okayama 714-1213}
\author{Patrick \textsc{Schmeer}}
\affil{Bischmisheim, Am Probstbaum 10, 66132 Saarbr\"ucken, Germany}


\KeyWords{accretion, accretion disks
          --- stars: novae, cataclysmic variables
          --- stars: dwarf novae
          --- stars: individual (QW Ser)}

\maketitle

\begin{abstract}

 We report on the results of the QW Ser campaign which has been continued
 from 2000 to 2003 by the VSNET collaboration team.  Four long outbursts
 and many short ones were caught during this period.  Our intensive
 photometric observations revealed superhumps with a period of 0.07700(4)
 d during all four superoutbursts, proving the SU UMa nature of this
 star.  The recurrence cycles of the normal outbursts and the
 superoutbursts were measured to be $\sim$50 days and 240(30) days,
 respectively.  The change rate of the superhump period was
 $-5.8\times10^{-5}$.  The distance and the X-ray luminosity in the
 range of 0.5-2.4 keV are estimated to be 380(60) pc and $\log L_{\rm X}
 = 31.0 \pm 0.1$ erg s$^{-1}$.  These properties have typical values for
 an SU UMa-type dwarf nova with this superhump period.

\end{abstract}

\section{Introduction}

Cataclysmic variable stars (CVs) are a group of binary stars
consisting of a white dwarf (primary star) and a late-type secondary
star (for a thorough review, \cite{war95book}).  Surface gas of the
secondary pouring out from its Roche lobe via the inner Lagrange point
is transferred to the primary.  The gas is accreted by the white dwarf
through an accretion disk in usual CVs.

Dwarf novae are a class of CVs, repeatedly showing large amplitude
variations of brightness (2--8 mag) due to changes of the physical status
of the accretion disk.  SU UMa-type dwarf novae, which originally
defined by \citep{war85suuma}, give rise to two types of outbursts:
normal outbursts and long lasting superoutbursts.  Superhumps are
small-amplitude modulations characteristic to SU UMa stars which are
observed only during the superoutburst.  Prograde precession of the
eccentric accretion disk due to the tidal influence has been attributed
to the cause of the superhump phenomenon \citep{whi88tidal}.  The
thermal-tidal disk instability model is the currently standard model for
the dwarf nova-type outbursts, and has succeeded in reproducing the
basic properties of various types of the brightness variation in dwarf
novae including SU UMa stars (for a review, see \cite{osa96review}; for
an example of two-dimensinal simulations, \cite{tru01DNsuperoutburst}).
Observations of dwarf novae are, however, still producing problems which
challenge the 'current' model.

QW Ser was discovered by \citet{tak98qwser} who detected four positive
detections in his photographic film collection.  He designated this
variable star as TmzV46, and suggested it to be a possible dwarf nova,
based on its blue color in the USNO A1.0 catalog.  \citet{sch99qwser}
detected an outburst at 1999 Oct. 4.114 (UT), confirming the dwarf nova
classification.  Tracing this outburst, \citet{kat99qwser} revealed that
the outburst duration was between 11 d and 16 d and the decline rate was
0.10 mag d$^{-1}$.  \citet{NameList75} finally gave TmzV46 the permanent
variable star name, QW Ser.

QW Ser is identified with USNO B1.0 0983-0296263 ($B1 = 17.57, R1 =
17.44, B2 = 18.20, R2 = 17.42$).  The proper motion of this star is
listed as ($\mu_{\rm R.A.}, \mu_{\rm Dec}$) = ($-$4(2), $-$40(2)) in a
unit of mas yr$^{-1}$.  QW Ser is also identified with the X-ray source
1RXS J152613.9+081845 \citep{ROSATFSCiauc}, which has a 52--201 keV
count rate of 0.045(18) count s$^{-1}$.

We in this paper report on our observations during four long outbursts
in 2000, 2001, 2002, and 2003, which unveiled the SU UMa nature of QW
Ser.  The next section mentions the observations, and the section 3
describes the details of our observational results.  The characteristics
of QW Ser will be discussed in the section 4.

\begin{table*}
\caption{Log of observations.}\label{tab:log}
\begin{center}
\begin{tabular}{clrcccccc}
\hline\hline
\multicolumn{3}{c}{Date} & HJD-2400000 & Exposure & Frame  & Comp.
 & filter & Instrument$^{\dagger}$ \\
         & &             & Start--End & Time (s) & Number & Star$^*$  &
      &     \\
\hline
2000 & July &  6 & 51732.334--51732.516 & 30  & 297 & 5 & $R_{\rm c}$  & N   \\
     &      &    & 51732.347--51732.431 & 90  &  64 & 2 & $R_{\rm j}$  & P   \\
     &      &  7 & 51733.299--51733.394 & 60  &  60 & 2 & $R_{\rm j}$  & P   \\
     &      &  8 & 51733.952--51734.082 & 30  &  62 & 2 & $V$  & K   \\
     &      &    & 51734.411--51734.508 & 20  & 284 & 2 & no & Ma  \\
     &      &    & 51734.280--51734.301 & 60  &  21 & 2 & $R_{\rm j}$  & P   \\
     &      &  9 & 51735.077--51735.118 & 60  &  49 & 2 & $V$  & K   \\
     &      &    & 51735.354--51735.482 & 20  & 380 & 5 & no & Ma  \\
     &      & 10 & 51735.948--51736.087 & 60  & 164 & 2 & $V$  & K   \\
     &      & 11 & 51737.304--51737.339 & 60  &  35 & 2 & $V$  & P   \\
     &      &    & 51737.406--51737.498 & 30  & 232 & 5 & $R_{\rm c}$  & N   \\
     &      & 12 & 51737.970--51738.035 & 60  &  77 & 2 & $V$  & K   \\
     &      &    & 51738.296--51738.382 & 60  &  94 & 2 & $V$  & P   \\
     &      &    & 51738.346--51738.500 & 50  & 195 & 5 & $R_{\rm c}$  & N   \\
     &      & 13 & 51739.295--51739.380 & 100 &  61 & 2 & $V$  & P   \\
     &      &    & 51739.326--51739.509 & 50  & 230 & 5 & $R_{\rm c}$  & N   \\
     &      & 14 & 51740.284--51740.379 & 100 &  70 & 2 & $V$  & P   \\
     &      & 15 & 51740.717--51740.790 & 16  & 240 & 2 & no & C   \\
     &      &    & 51741.281--51741.352 & 100 &  46 & 2 & $V$  & P   \\
     &      & 16 & 51742.319--51742.394 & 60  &  85 & 2 & $R_{\rm j}$  & P   \\
     &      & 17 & 51743.330--51743.338 & 100 &   3 & 2 & $R_{\rm j}$  & P   \\
     &      & 18 & 51745.328--51745.381 & 100 &  37 & 2 & $R_{\rm j}$  & P   \\
     &      & 19 & 51746.292--51746.304 & 200 &   3 & 2 & $R_{\rm j}$  & P   \\
     &      & 20 & 51747.277--51747.315 & 100 &  21 & 2 & $R_{\rm j}$  & P   \\
     &      & 21 & 51748.356--51748.357 & 100 &   1 & 2 & $R_{\rm j}$  & P   \\
     &      & 23 & 51750.326--51750.346 & 200 &   9 & 2 & $R_{\rm j}$  & P   \\
     &      & 24 & 51751.275--51751.285 & 200 &   4 & 2 & $R_{\rm j}$  & P   \\
     &      & 25 & 51752.287--51752.302 & 200 &   5 & 2 & $R_{\rm j}$  & P   \\
     &      & 26 & 51753.287--51753.310 & 200 &   8 & 2 & $R_{\rm j}$  & P   \\
     &      & 27 & 51754.268--51754.283 & 200 &   7 & 2 & $R_{\rm j}$  & P   \\
     &      & 28 & 51755.269--51755.293 & 200 &  10 & 2 & $R_{\rm j}$  & P   \\
     &      & 29 & 51756.271--51756.279 & 200 &   4 & 2 & $R_{\rm j}$  & P   \\
2001 & Feb. & 10 & 51951.232--51951.349 & 60  & 135 & 2 & $V$  & K   \\
     &      & 11 & 51951.478--51951.725 & 60  & 232 & 2 & $R_{\rm c}$  & N   \\
     &      & 12 & 51952.537--51952.764 & 45  & 232 & 6 & $R_{\rm c}$  & N   \\
2002 & June &  2 & 52428.078--52428.257 & 30  & 340 & 1 & no & O25 \\
     &      &    & 52428.102--52428.180 & 15  & 254 & 1 & no & T   \\
     &      &  3 & 52428.999--52429.227 & 14  & 760 & 2 & no & M   \\
     &      &    & 52429.054--52429.174 & 30  & 255 & 1 & no & O25 \\
     &      &    & 52429.092--52429.243 & 15  & 628 & 1 & no & T   \\
     &      &    & 52429.140--52429.284 & 30  & 331 & 1 & no & O30 \\
     &      &  4 & 52430.087--52430.214 & 30  & 239 & 1 & no & O30 \\
     &      &    & 52430.128--52430.236 & 30  & 193 & 1 & no & O25 \\
     &      &  5 & 52431.036--52431.178 & 10  & 1450& 1 & no & T   \\
     &      &  6 & 52431.960--52432.059 & 30  & 200 & 3 & $V$  & K   \\
     &      &    & 52431.991--52432.260 & 14  & 904 & 2 & no & M   \\
     &      &    & 52432.038--52432.238 & 30  & 475 & 1 & no & O25 \\
     &      &  7 & 52432.998--52433.074 & 30  & 175 & 3 & $V$  & K   \\
     &      &  9 & 52435.028--52435.170 & 10  & 539 & 4 & no & H   \\
     &      &    & 52435.058--52435.236 & 10  & 871 & 1 & no & T   \\
     &      & 15 & 52441.020--52441.027 & 10  &  31 & 1 & no & O30 \\
     &      &    & 51441.334--52441.335 & 120 &   1 & 2 & $R_{\rm j}$  & P   \\
     &      & 16 & 52441.984--52442.000 & 30  &  29 & 1 & no & O25 \\
     &      &    & 52442.326--52442.331 & 120 &   2 & 2 & $R_{\rm j}$  & P   \\
\hline
 \end{tabular}
\end{center}
\end{table*}

\setcounter{table}{0}
\begin{table*}
\caption{(continued)}
\begin{center}
\begin{tabular}{clrcccccc}
\hline\hline
\multicolumn{3}{c}{Date} & HJD-2400000 & Exposure & Frame  & Comp.
 & filter & Instrument$^{\dagger}$ \\
         & &             & Start--End & Time (s) & Number & Star$^*$  &
      &     \\
\hline
2002 & June & 20 & 52446.315--52446.322 & 120 &   4 & 2 & $R_{\rm j}$  & P   \\
     &      & 21 & 52447.311--52447.309 &  60 &   6 & 2 & $R_{\rm j}$  & P   \\
     &      & 22 & 52448.310--52448.317 & 120 &   5 & 2 & $R_{\rm j}$  & P   \\
     &      & 23 & 52449.307--52449.312 & 120 &   3 & 2 & $R_{\rm j}$  & P   \\
     &      & 25 & 52452.308--52452.306 & 240 &   3 & 2 & $R_{\rm j}$  & P   \\
     &      & 29 & 52456.330--52456.333 & 120 &   2 & 2 & $R_{\rm j}$  & P   \\
     & July &  2 & 52457.983--52457.990 &  30 &  15 & 1 & no & O30 \\
2003 & Feb. & 24 & 52695.151--52695.375 &  30 & 214 & 1 & no & O25 \\
     &      & 25 & 52696.193--52696.372 &  30 & 325 & 1 & no & O25 \\
     &      &    & 52696.169--52696.220 &  30 & 111 & 2 & no & H   \\
     &      & 27 & 52698.181--52698.346 &  30 & 307 & 1 & no & O25 \\
     &      &    & 52698.192--52698.294 &  40 & 214 & 2 & no & K   \\
     &      & 28 & 52699.135--52699.147 &  30 &  22 & 1 & no & O25 \\
     & Mar. &  5 & 52704.153--52704.245 &  30 &  98 & 1 & no & O25 \\
\hline
\multicolumn{9}{l}{$^*$Comparison star 1: HD 137532 (a close double star
 of combined $V\sim9.7$, noted in the Henden\&Sumner}\\
\multicolumn{9}{l}{(H\&S) sequence, 2: $V$=13.411(6) and $B-V$=0.673(4)
 in the H\&S sequence (ID 4), 3: $V$=11.977(8)}\\
\multicolumn{9}{l}{and $B-V$=0.946(8) (ID 2), 4:
 $V$=11.770(1) and $B-V$=0.637(4) (ID 1), 5: $V$=13.120(12) and}\\
\multicolumn{9}{l}{$B-V$=0.722(13) (ID 3), 6:
 $V$=14.599(0) and $B-V$=0.638(5) (ID 7)}\\
\multicolumn{9}{l}{$^\dagger$Instrument N: 40-cm telescope + ST-7 (Brno,
 Czech), P: 38-cm Telescope + SBIG ST-7 (Crimea,}\\
\multicolumn{9}{l}{Ukraine), K: 25-cm telescope + Apogee AP-7 (Tsukuba,
 Japan), Ma: 28-cm telescope + SBIG ST-7}\\
\multicolumn{9}{l}{(Ceccano, Italy), C: 44-cm telescope + Genesis 16\#90
 (KAF 1602e) (California, USA), O25: 25-cm}\\
\multicolumn{9}{l}{telescope + SBIG ST-7/ST-7E (Kyoto, Japan), T: 30-cm
 telescope + SBIG ST-9E (Okayama, Japan),}\\
\multicolumn{9}{l}{O30: 30-cm telescope + SBIG ST-7/ST-7E (Kyoto,
 Japan), M: 25-cm telescope + SBIG ST-7}\\
\multicolumn{9}{l}{(Okayama, Japan), H: 60-cm telescope + PixCellent
 S/T 00-3194 (SITe 003AB) (Hida, Japan)}

\end{tabular}
\end{center}
\end{table*}

\section{Observation}

\begin{figure}
  \begin{center}
    \FigureFile(84mm,84mm){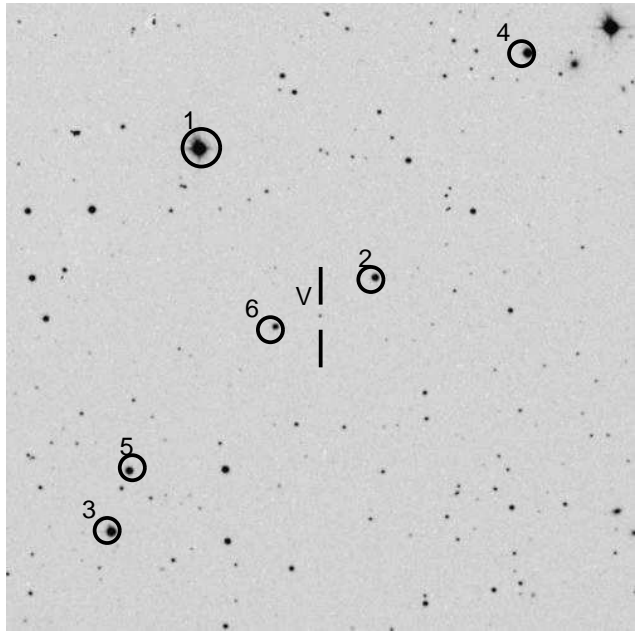}
  \end{center}
 \caption{Finding chart of QW Ser generated by the astronomical
 image-data server operated by the National Astronomical Observatory of
 Japan, making use of Digital Sky Survey 2.  North is up, and East is
 left.  The comparison stars used are given their numbers, which are
 identical with those in Table 1.
 }
  \label{fig:chart}
\end{figure}

The observations were carried out at nine sites with ten sets of
instruments.  The log of the observations and the instruments are
summarized in Table \ref{tab:log}.  Figure \ref{fig:chart} is a finding
chart where the local comparison stars used are marked.

The Kyoto and Okayama frames were processed by the PSF photometry
package developed by one of the authors (TK) after dark-subtraction and
flat-fielding.  
All the frames obtained at Hida were reduced by the aperture photometry
package in IRAF\footnote{IRAF is distributed by the National Optical
Astronomy Observatories for Research in Astronomy, Inc. under
cooperative agreement with the National Science Foundation.}, after
de-biasing and flat-fielding.  All frames obtained at the CBA Concord
and Rome were reduced by aperture photometry after dark subtraction and
flat-fielding, using the AIP4WIN software by Berry and
Burnell\footnote{$\langle$http://www.willbell.com/aip/index.htm$\rangle$}
and the QMiPS32 software, respectively.  The Crimean images were
dark-subtracted, flat-fielded, and analyzed with the profile/aperture
photometry package developed by Vitalij~P.~Goranskij.  PSF photometry of
the Brno data were performed, using the package
Munidos\footnote{$\langle$http://munipack.astronomy.cz$\rangle$} which
is based on Daophot II.

The magnitude scale was calibrated using the Henden\&Sumner
sequence,\footnote{
$\langle$ftp://ftp.nofs.navy.mil/pub/outgoing/aah/sequence/sumner/qwser.seq$\rangle$}
and all the data were adjusted to match the $V$-band data obtained at
Tsukuba and Crimea.  The heliocentric correction was applied to the
observation times before the following analyses.

\section{Result}

The campaigns of time-resolved photometry were conducted during four
superoutbursts which occurred in 2000 July, 2001 February, 2002 June,
and 2003 February.  In this section, we describe the detailed results of
four campaigns separately.

\subsection{The 2000 July superoutburst}

The first superoutburst was noticed by P. Schmeer at 2000 July 5.941
(UT).  We started observations on July 6, the day when we received the
outburst report.

\begin{figure}
  \begin{center}
    \FigureFile(84mm,115mm){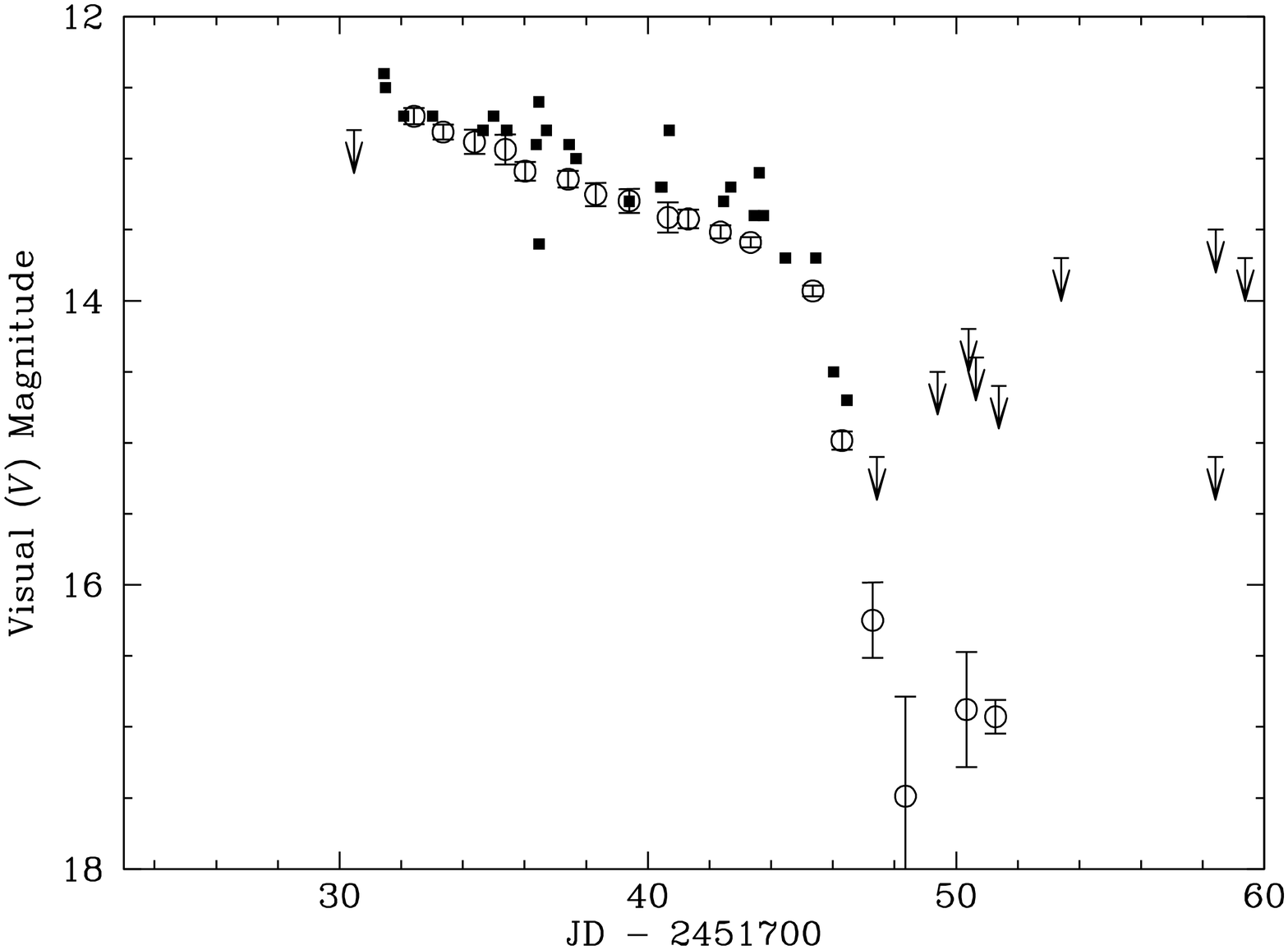}
  \end{center}
 \caption{Long-term light curve of the 2000 superoutburst.  The filled
 squares and down arrows are the visual (or CCD) observations and the
 upper limits reported to VSNET.  The abscissa is $Julian Day - 2451700$,
 and the ordinate is $Visual (V) magnitude$.  The open circles with the
 error bars indicate the daily mean $V$ magnitudes of our data and their
 standard deviations.
 } \label{fig:0007long}
\end{figure}

The long-term light curve of the 2000 July superoutburst is drawn in
figure \ref{fig:0007long}.  The superoutburst lasted for 15 days, and QW
Ser declined by an almost constant rate of 0.086 mag d$^{-1}$ during the
plateau phase.  Then, the variable entered the rapid decline phase of a
rate of 1.2 mag d$^{-1}$.  While we could not detect a rebrightening,
QW Ser seemed to remain about $V = 17.0$ for at least a few days after
the superoutburst.

\begin{figure}
  \begin{center}
    \FigureFile(84mm,115mm){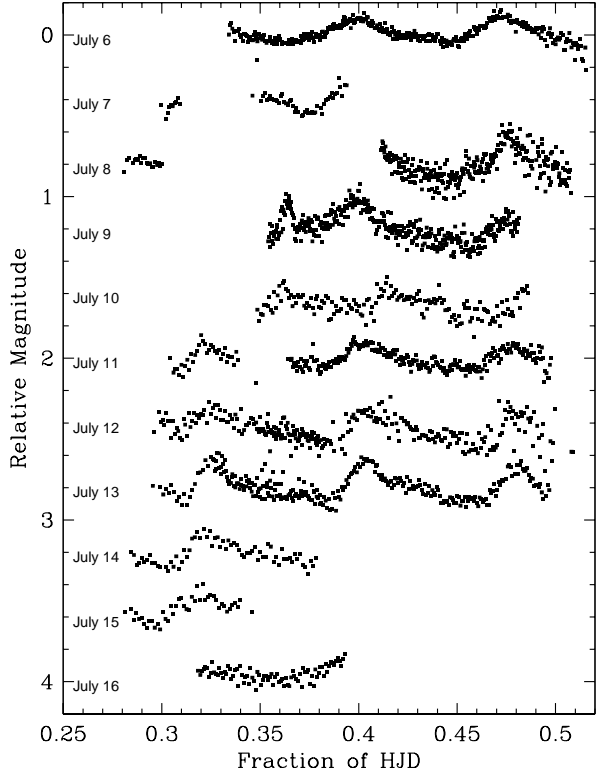}
  \end{center}
 \caption{Daily light curves of the 2000 superoutburst during the
 plateau phase.  Each daily dataset is shifted by +0.4 mag.  The
 superhump had already grown to 0.16 mag by July 6, within 2 days from
 the superoutburst onset.  The superhump amplitude increased to 0.35 mag
 by July 8, then decreased.  However, it again increased to 0.26 mag by
 July 13, then seemed to decrease again.  On July 9, we can see a flare
 of 0.25 mag with a time scale of $\sim$15 min beside a superhump
 maximum around a fraction of HJD of 0.4.
 } \label{fig:0007shshape}
\end{figure}

Daily light curves are shown in figure \ref{fig:0007shshape}.  The data
obtained on 2000 July 6, the first night of our observations, clearly
shows superhumps with an amplitude of 0.16 mag, which first proved the
SU UMa nature of this dwarf nova.  The superhumps had developed within 2
days from the onset of the outburst (figure \ref{fig:0007long}).  The
amplitude of the superhumps continued to increase to 0.35 mag by July 8,
and then started to decline.  The amplitude, however, again developed to
0.26 mag by July 13, and seemed to decline, suggesting the timescale of
the variation of the superhump amplitude to be 5 days.  The rise of the
superhump gradually but steadily became steeper by July 13, and seemed
to get more gradual after that.  We can not see a sign of emergence of
the secondary hump until July 14, the 9th day of the superoutburst.

\begin{figure}
  \begin{center}
    \FigureFile(84mm,115mm){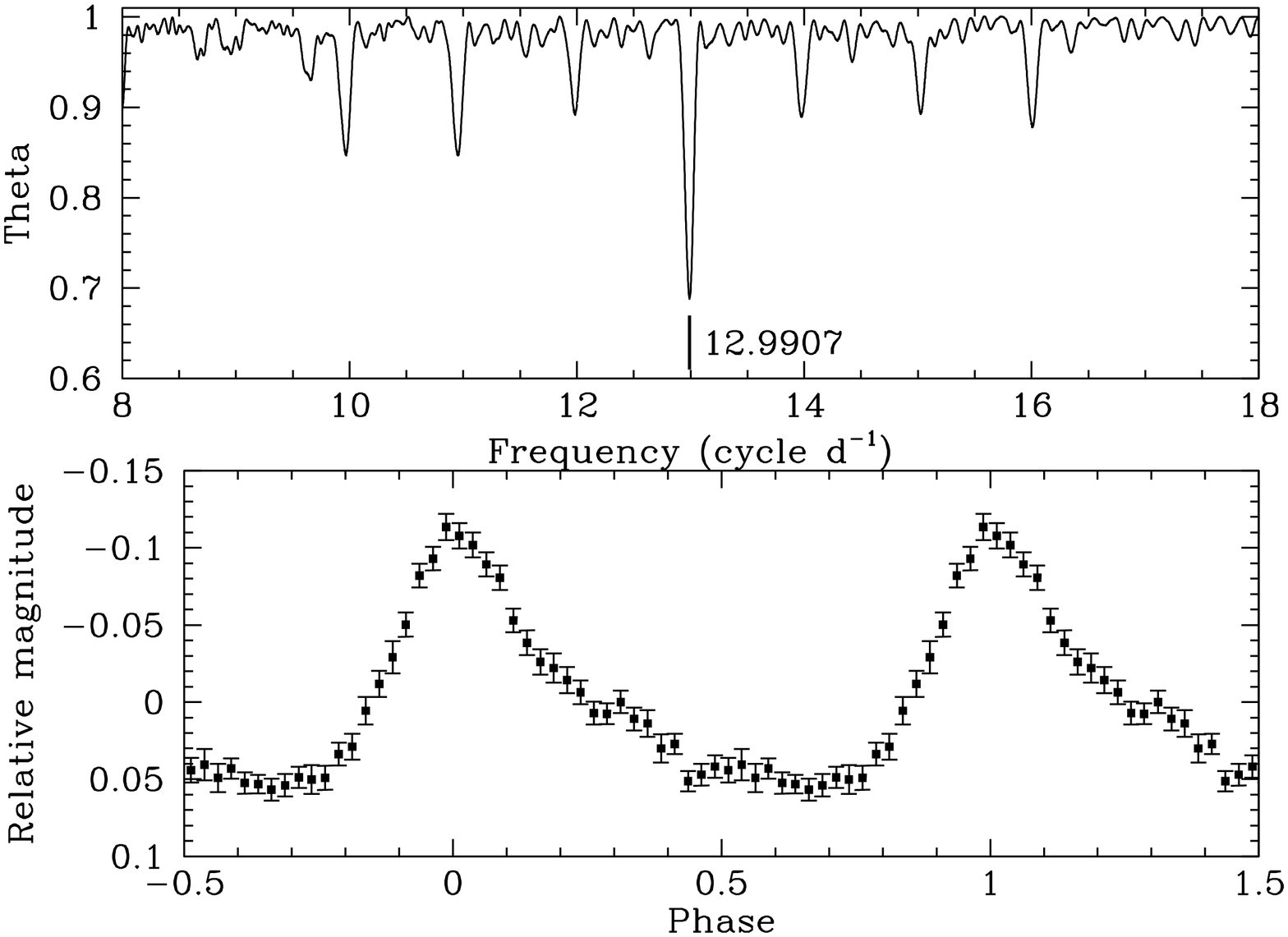}
  \end{center}
 \caption{(upper panel) Theta diagram obtained by a PDM period analysis
 for the data obtained in July 6--17.  The best estimated superhump
 period is 0.076978(8) d (12.9908(13) cycle d$^{-1}$).  (lower panel)
 The superhump light curve folded by that period.
 } \label{fig:0007sh}
\end{figure}

In the light curve of July 9, a 0.25-mag flare with a timescale of about
15 min was present before the superhump maximum around a fraction of HJD
of 0.4.  The light curve of a local check star relative to the
comparison star did not show a special feature around that time.

After subtraction of the linear decline trend, we performed the period
analysis using the Phase Dispersion Minimization (PDM) method
\citep{PDM} for the data obtained in July 6--17.  The resultant $\Theta$
diagram (figure \ref{fig:0007sh}) indicates 0.076978(8) d
($f=12.9907(14)$ cycle d$^{-1}$) to be the best estimated average
superhump period ($P_{\rm SH}$).  The error of the period was estimated
using the Lafler-Kinman class of methods, as applied by
\citet{fer89error}.  The lower panel of figure \ref{fig:0007sh} displays
the average superhump light curve which was yielded by folding the
de-trended light curves by the superhump period.

\begin{figure}
  \begin{center}
    \FigureFile(84mm,115mm){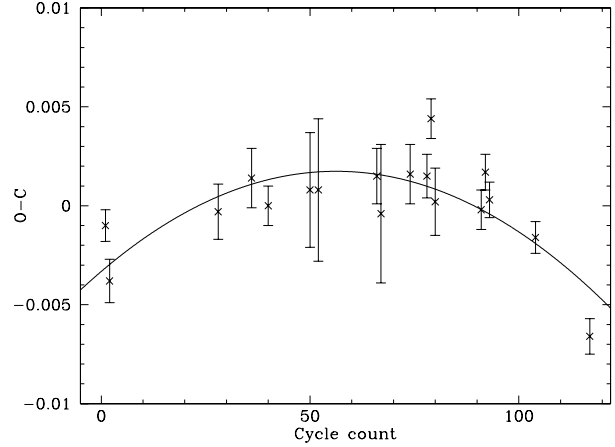}
  \end{center}
 \caption{O$-$C diagram of the superhump maximum timings during the 2000
 superoutburst.  The data is listed in the column O$-$C1 in table
 \ref{tab:0007max}.  The solid curve represents the quadratic polynomial
 obtained by fittng the O$-$C values (equation (2)), showing that the
 superhump period decreased with a rate of $\dot{P}_{\rm SH}/P_{\rm SH} =
 -4.2(0.8)\times 10^{-5}$.
 } \label{fig:0007max}
\end{figure}

The timings of the superhump maxima were extracted by fitting the average
superhump light curve in figure \ref{fig:0007sh}.  Table
\ref{tab:0007max} lists the results.  The cycle count was set to 1 at
the first superhump maximum observed on July 6.  Linear regression of
the superhump maximum timings yields the following equation:
\begin{equation}
 HJD_{\rm max} = 32.3210(13) + 0.076963(18)\times E.
\end{equation}
Figure \ref{fig:0007max} and table \ref{tab:0007max} show the results
of subtraction of the calculated maximum timings by equation (1) from
the observed ones (O$-$C1).  The values of O$-$C1 are fitted to the
following quadratic polynomial:
\begin{eqnarray}
 O-C1 = & 0.0017(5) - 1.2(1.2) \times 10^{-5}(E-60) \nonumber \\
        & - 1.6(0.3) \times 10^{-6}(E-60)^2 .
\end{eqnarray}
This curve is also drawn in figure \ref{fig:0007max}.  The derived index
of the quadratic term means that the superhump period decreased with a
rate of $\dot{P}_{\rm SH}/P_{\rm SH} = -4.2(0.8)\times 10^{-5}$.  Note
that the superhump period seemed constant in $E$ = 35--95.

\begin{table}
\caption{Timings of the superhump maxima during the 2000 July
 superoutburst.}\label{tab:0007max}
\begin{center}
\begin{tabular}{crrr}
\hline\hline
HJD$-$2451700 &  E & O$-$C1$^*$ &  O$-$C2$^\dagger$ \\
\hline
32.3969(08) &   1 & $-$0.0010 &  0.0021 \\
32.4711(11) &   2 & $-$0.0038 & $-$0.0008 \\
34.4756(14) &  28 & $-$0.0003 & $-$0.0008 \\
35.0930(15) &  36 &  0.0014 &  0.0003 \\
35.3995(10) &  40 &  0.0000 & $-$0.0013 \\
36.1699(29) &  50 &  0.0008 & $-$0.0009 \\
36.3238(36) &  52 &  0.0008 & $-$0.0009 \\
37.4020(14) &  66 &  0.0015 & $-$0.0001 \\
37.4771(35) &  67 & $-$0.0004 & $-$0.0020 \\
38.0178(15) &  74 &  0.0016 &  0.0004 \\
38.3256(11) &  78 &  0.0015 &  0.0005 \\
38.4054(10) &  79 &  0.0044 &  0.0035 \\
38.4782(17) &  80 &  0.0002 & $-$0.0006 \\
39.3244(10) &  91 & $-$0.0002 & $-$0.0000 \\
39.4033(09) &  92 &  0.0017 &  0.0020 \\
39.4788(09) &  93 &  0.0003 &  0.0007 \\
40.3235(08) & 104 & $-$0.0016 &  0.0003 \\
41.3190(09) & 117 & $-$0.0066 & $-$0.0024 \\ \hline
\multicolumn{4}{l}{$^*$ Using equation (1).}\\
\multicolumn{4}{l}{$^\dagger$ Using equation (2).}\\
\end{tabular}
\end{center}
\end{table}

\subsection{The 2001 February superoutburst}

\begin{figure}
  \begin{center}
    \FigureFile(84mm,115mm){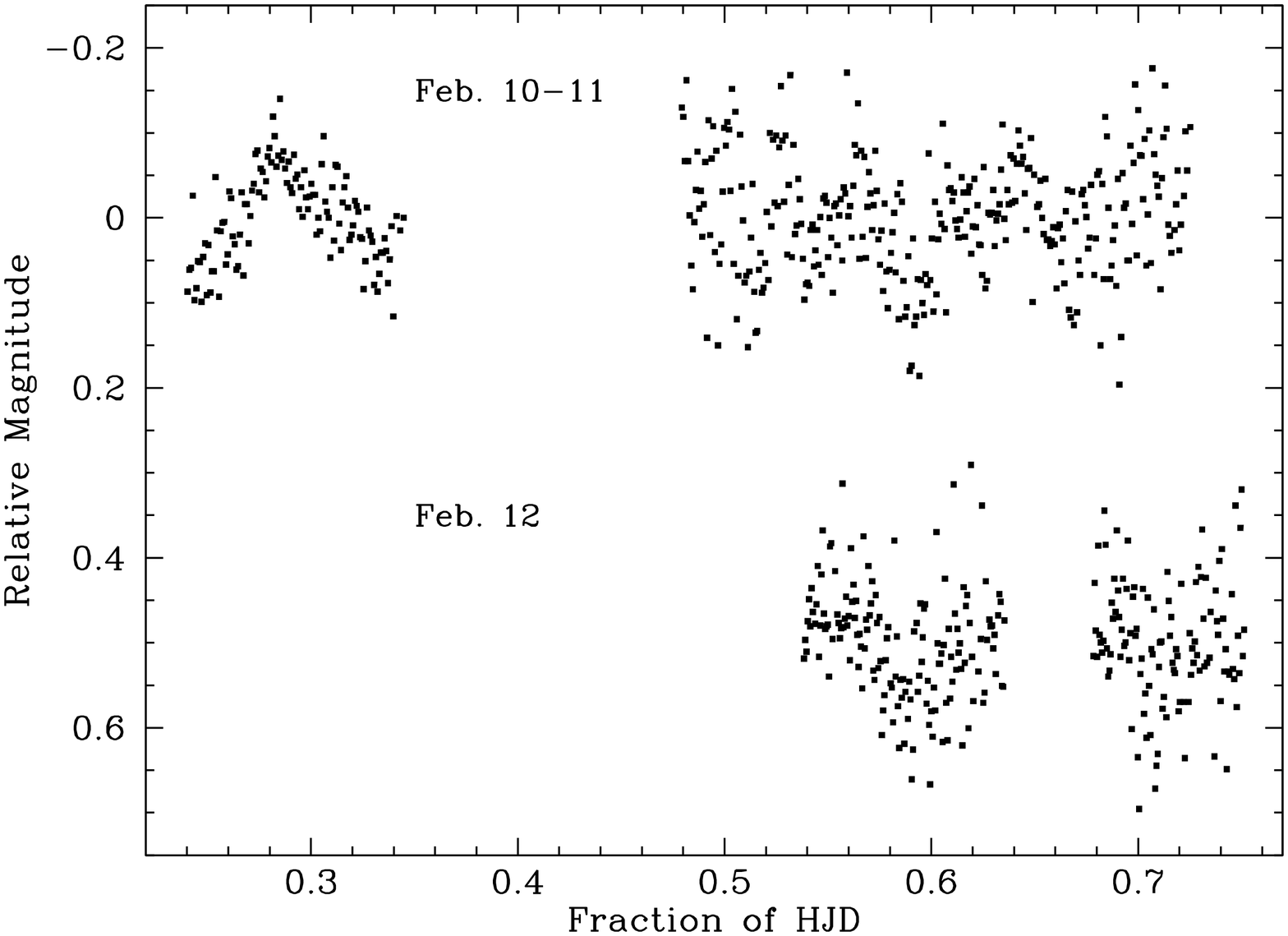}
  \end{center}
 \caption{Short-term light curves of the 2001 February outburst.
 Each daily dataset is shifted by +0.5 mag.  Although the errors were
 not  so small, superhumps certainly existed, proving that this outburst
 is a  genuine superoutburst.
 } \label{fig:0102shshape}
\end{figure}

The data available via the VSNET data browser\footnote{$\langle$
http://vsnet.kusastro.kyoto-u.ac.jp/vsnet/etc/searchobs.html$\rangle$}
around the 2001 February superoutburst are 13.2 mag at 2001 January
25.229 (UT), 16.9 mag at 30.86 (UT), and 13.5 mag at February 12.46
(UT).  No upper-limit was reported to VSNET during this period.  We
carried out time-resolved photometry on February 10, 11, and 12.  All
the data are represented in figure \ref{fig:0102shshape}.  We can see
obvious superhumps above the noise level of each dataset.  Thus this
outburst was surely a superoutburst.  The positive detection on January
25 was probably a precursor of this superoutburst.

\subsection{The 2002 June superoutburst}

\begin{figure}
  \begin{center}
    \FigureFile(84mm,115mm){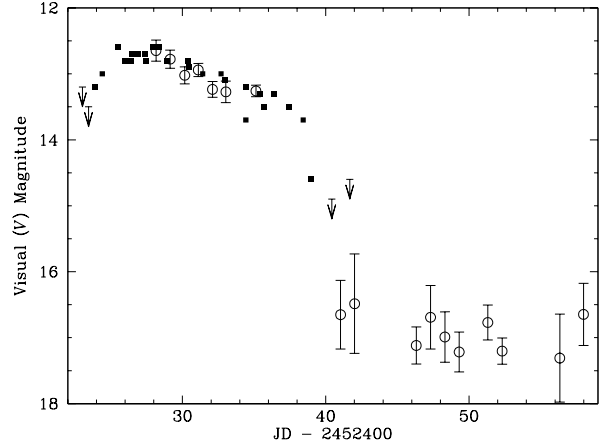}
  \end{center}
 \caption{Long-term light curve of the 2002 superoutburst, as figure
 \ref{fig:0007long}.
 } \label{fig:0206long}
\end{figure}

This outburst was caught by Rod Stubbings at 2002 May 29.410 (UT) at
$m_{\rm vis}=13.2$.  We started follow-up observations 5 days later. The
long-term light curve is presented in figure \ref{fig:0206long}.  QW Ser
faded with a rate of 0.13 mag d$^{-1}$ between HJD 2452428 and 2452433,
but appeared to remain at the same brightness between HJD 2452433 and
2452435.  After the superoutburst lasted 14 days, QW Ser remained about
$V \sim 17.0$ for at least two weeks.

\begin{figure}
  \begin{center}
    \FigureFile(84mm,115mm){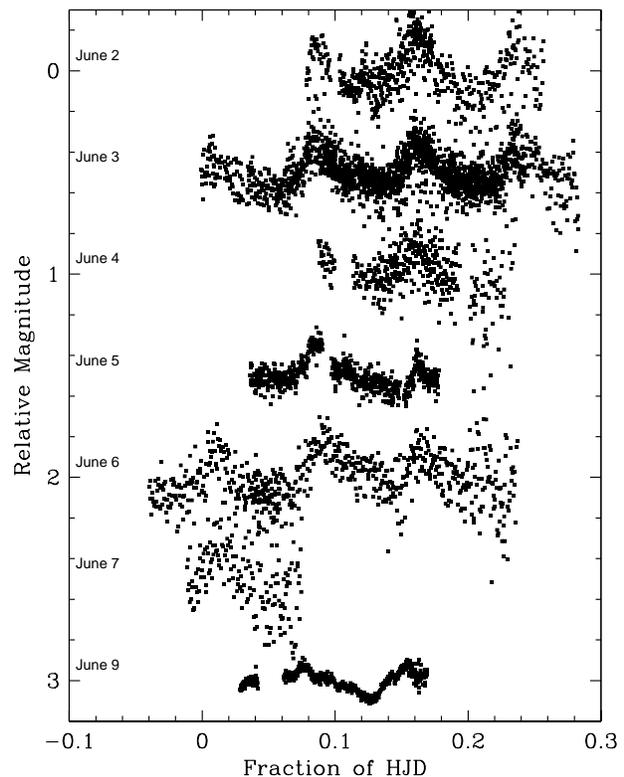}
  \end{center}
 \caption{Short-term light curves of the 2002 June outburst during
 the plateau phase.  Each daily dataset is shifted by +0.5 mag.
 } \label{fig:0206shshape}
\end{figure}

\begin{figure}
  \begin{center}
    \FigureFile(84mm,115mm){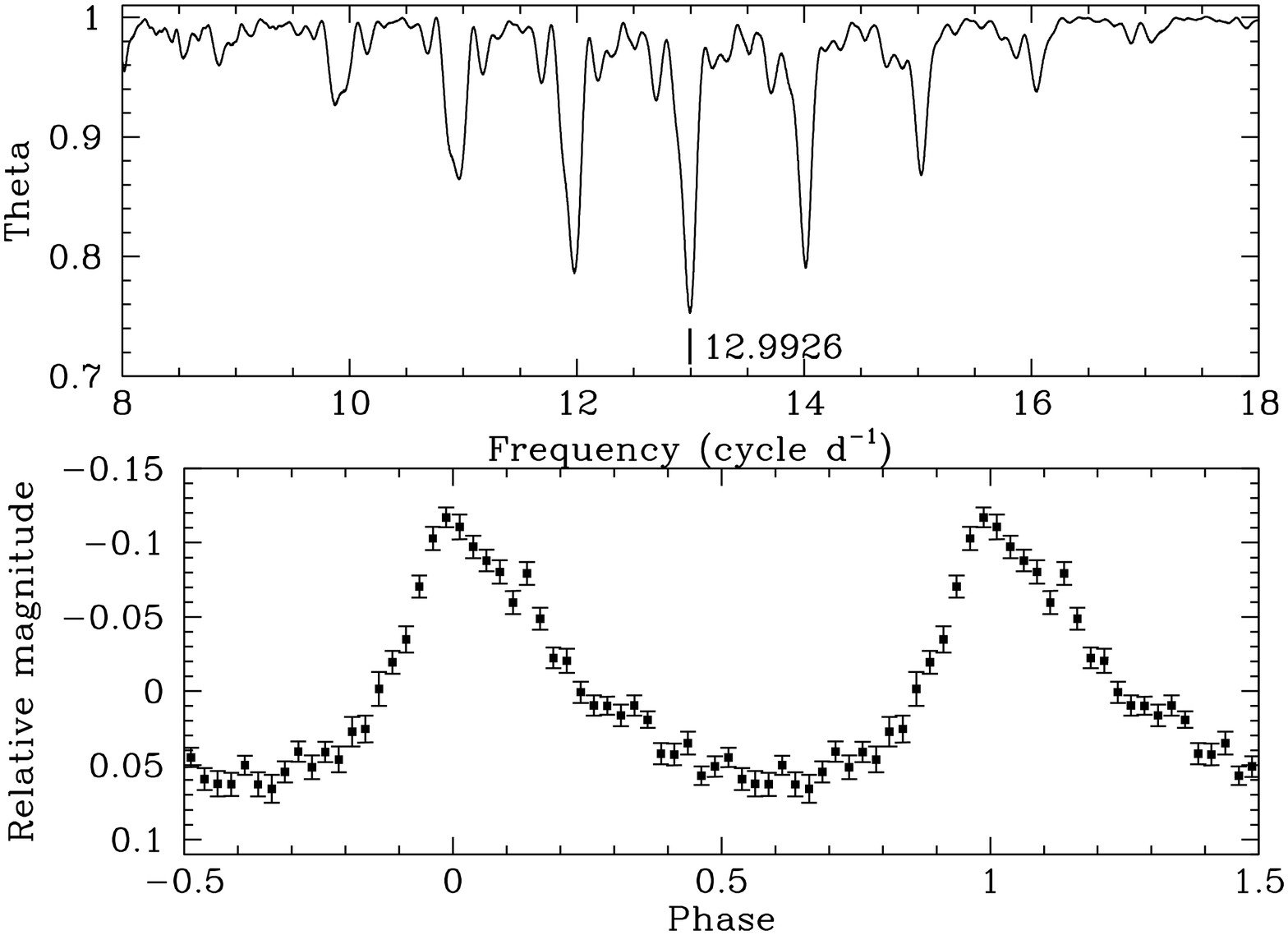}
  \end{center}
 \caption{(upper panel) Theta diagram obtained by a PDM period analysis
 for the data obtained in 2002 July 2--9.  The best estimated superhump
 period is 0.076967(13) d (12.9926(22) cycle d$^{-1}$).  (lower panel)
 The superhump light curve folded by that period.
 } \label{fig:0206sh}
\end{figure}

The superhumps were caught in all the dataset of each night (figure
\ref{fig:0206shshape}). We subtracted a linear decline trend of 0.13 mag
d$^{-1}$ from the plateau-phase data, and corrected the second-order
color effect of each night run.  After this pre-whitening, we analyzed
the data by the PDM method.  The resultant $\Theta$ diagram and the
averaged superhump light curve are in figure \ref{fig:0206sh}.  The best
estimated superhump period of 0.076967(13) d is equal to that in the
2000 superoutburst within the statistical error.

\begin{figure}
  \begin{center}
    \FigureFile(84mm,115mm){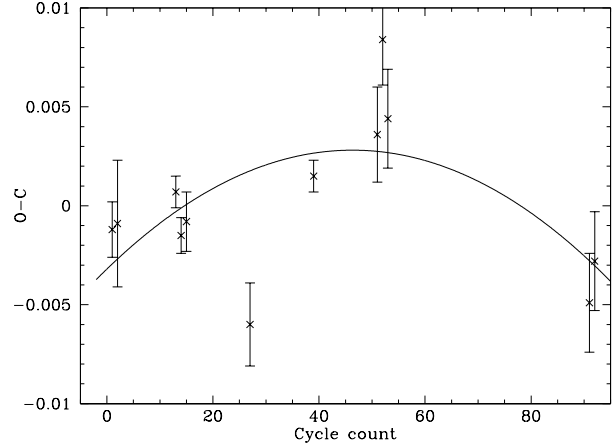}
  \end{center}
 \caption{O$-$C diagram of the superhump maximum timings during the 2002
 superoutburst.  The data is listed in the column O$-$C3 in table
 \ref{tab:0206max}.  The solid curve represents the quadratic polynomial
 obtained by fittng the O$-$C values (equation (4)), showing that the
 superhump period decreased with a rate of $\dot{P}_{\rm SH}/P_{\rm SH} =
 -7.3(3.1)\times 10^{-5}$.
 } \label{fig:0206max}
\end{figure}

The superhump maximum timings were measured in the way for the 2000
superoutburst, and listed in Table \ref{tab:0206max}.  The cycle count
was set to 1 at the first superhump maximum observed on June 2.  The
equation deduced by linear regression of the superhump maximum timings
is
\begin{equation}
 HJD_{\rm max} = 28.1598(13) + 0.076900(41)\times E.
\end{equation}
Figure \ref{fig:0206max} and table \ref{tab:0206max} show the results
of subtraction of the calculated maximum timings by equation (3) from
the observed ones (O$-$C3).  Fitting of the O$-$C3 by a quadratic
polynomial converges to
\begin{eqnarray}
 O-C3 = & 0.0028(16) + 0.2(3.3) \times 10^{-5}(E-46) \nonumber \\
        & - 2.8(1.2) \times 10^{-6}(E-46)^2 .
\end{eqnarray}
This curve is also drawn in figure \ref{fig:0206max}.  The $P_{\rm SH}$
decrease rate calculated from the index of the quadratic term in this
equation is $\dot{P}_{\rm SH}/P_{\rm SH} = -7.3(3.1)\times 10^{-5}$.  This
rate is negative, which is indicative of the $P_{\rm SH}$ decrease, and
consistent with that observed during the 2000 superoutburst within the
error.

\begin{table}
\caption{Timings of the superhump maxima during the 2002 June
 superoutburst.}\label{tab:0206max}
\begin{center}
\begin{tabular}{crrr}
\hline\hline
HJD$-$2452400 &  E & O$-$C3$^*$ &  O$-$C4$^\dagger$ \\
\hline
28.1587(14) &   1 & $-$0.0012 &  0.0017 \\
28.3459(32) &   2 & $-$0.0009 &  0.0018 \\
29.0834(08) &  13 &  0.0007   &  0.0010 \\
29.1581(09) &  14 & $-$0.0015 & $-$0.0014 \\
29.2357(15) &  15 & $-$0.0008 & $-$0.0009 \\
30.1533(21) &  27 & $-$0.0060 & $-$0.0078 \\
31.0836(08) &  39 &  0.0015   & $-$0.0012 \\
32.0085(24) &  51 &  0.0036   &  0.0009 \\
32.0902(23) &  52 &  0.0084   &  0.0057 \\
32.1631(25) &  53 &  0.0044   &  0.0017 \\
35.0760(25) &  91 & $-$0.0049 & $-$0.0021 \\
35.1550(25) &  92 & $-$0.0028 &  0.0002 \\ \hline
\multicolumn{4}{l}{$^*$ Using equation (3).}\\
\multicolumn{4}{l}{$^\dagger$ Using equation (4).}\\
\end{tabular}
\end{center}
\end{table}

\subsection{The 2003 February superoutburst}

Following the outburst detection ($m_{\rm vis}=12.6$) by E. Muyllaert at
2003 February 23.215 (UT), we started photometric observations on
February 24.

\begin{figure}
  \begin{center}
    \FigureFile(84mm,115mm){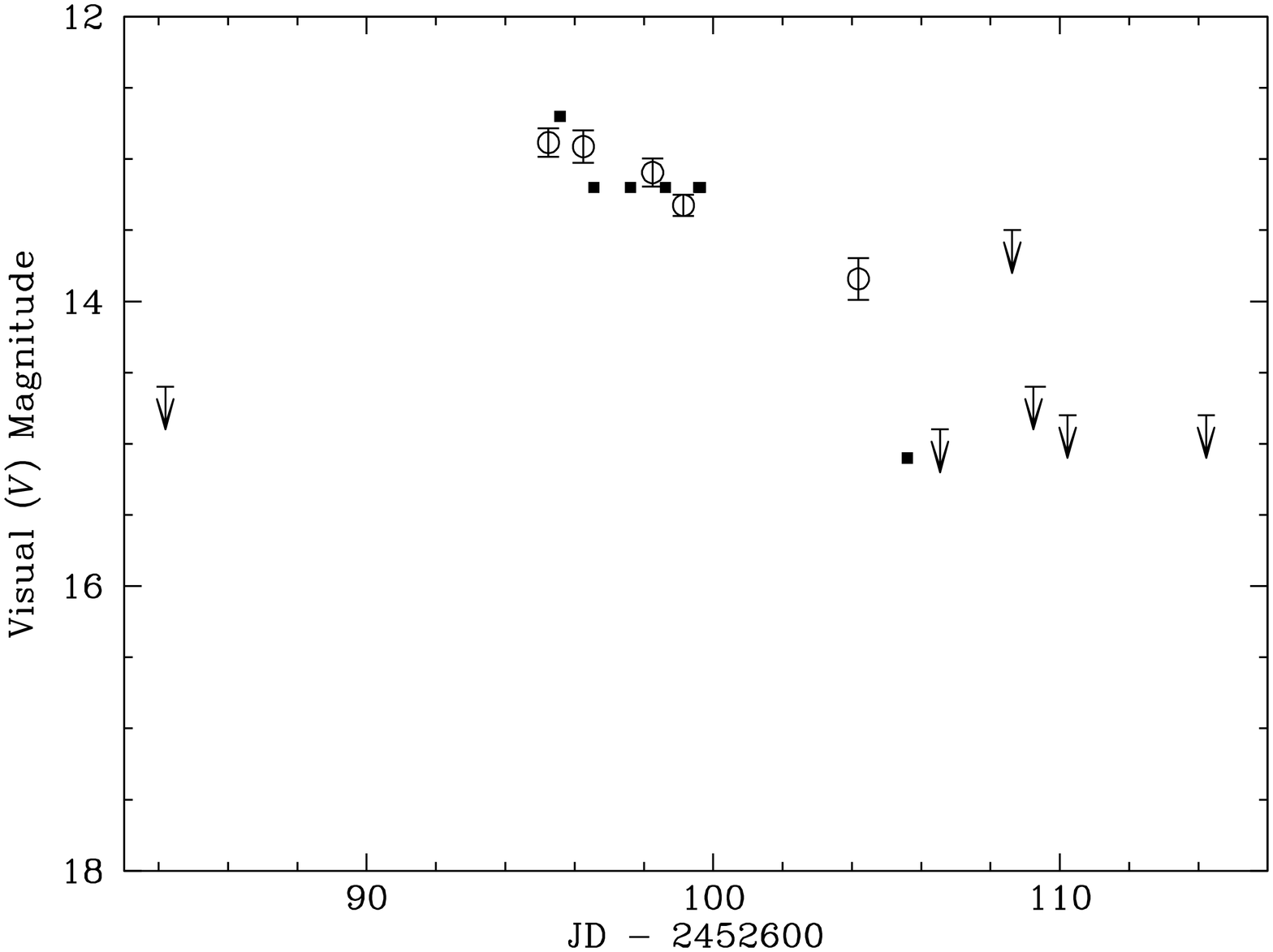}
  \end{center}
 \caption{Long-term light curve of the 2003 superoutburst, as figure
 \ref{fig:0007long}.
 } \label{fig:0302long}
\end{figure}

Figure \ref{fig:0302long} exhibits the long-term light curve of the 2003
July superoutburst.  The decline rate derived our data was 0.11 mag
d$^{-1}$.   Then, the variable entered the rapid decline phase of a
rate of 1.2 mag d$^{-1}$.

\begin{figure}
  \begin{center}
    \FigureFile(84mm,115mm){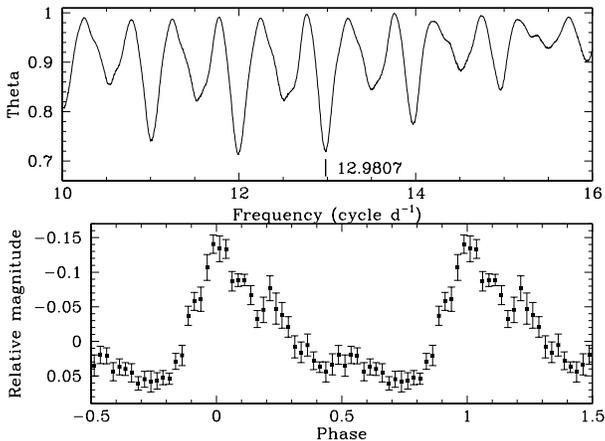}
  \end{center}
 \caption{Theta diagram for the data obtained in February 24--27 (upper
 panel), and the averaged superhump light curve (lower panel), as figure
 \ref{fig:0007sh}.  Due to the short coverages of each run, the periodgram
 suffers from an alias problem.  However, we can safely choose
 0.07704(4) d (12.981(6) cycle d$^{-1}$) as the best estimated $P_{\rm
 SH}$, because of those in the other superoutbursts.
 } \label{fig:0302sh}
\end{figure}

After prewhitening in the same way as for the data of the 2002
superoutburst, we performed the period analysis using the PDM method for
the data obtained on February 24, 25, and 27 (figure \ref{fig:0302sh}).
Although the short coverages of each dataset hinder us from
distinguishing the true signal from its aliases, we can safely choose the
genuine $P_{\rm SH}$ of 0.077037(37) d ($f=12.9807(62)$ cycle d$^{-1}$)
from the superhump periods during the other superoutbursts.  The lower
panel of figure \ref{fig:0302sh} presents the average superhump light
curve during our observations of this superoutburst.

The superhump maximum timings observed were insufficient to
significantly deduce the change rate of $P_{\rm SH}$.

\section{Discussion}

Table \ref{tab:outburst} summarizes the outbursts reported in
\cite{tak98qwser} and to VSNET and those detected by the All Sky
Automated Survey \citep{poj02ASAS}.  As the maximum magnitude of the
1999 superoutburst (12.2 mag) is suspected to be due to inaccuracy of
magnitudes of comparison stars in a finding chart the observer used, the
true superoutburst maximum should be around $V=12.5$.  Thus the
superoutburst amplitude is $\sim$5.0 mag.  The maximum magnitude of the
normal outburst is $V\sim13.1$.  The recurrence cycle of the normal
outburst seems to have been rather stable around 50 days.  If we assume
that a superoutburst was missed around 2001 September, the recurrence
cycle of the superoutburst (supercycle) has been also stable, about
220--270 d since 1999, while SU UMa stars showing variable outburst
patterns have recently been discovered, such as MN Dra
\citep{nog03var73dra}, DI UMa \citep{fri99diuma}, SU UMa
\citep{ros00suuma, kat02suuma}, V1113 Cyg \citep{kat01v1113cyg}, V503
Cyg \citep{kat02v503cyg}, and DM Lyr \citep{nog03dmlyr}.  The change
rate of the superhump period was $-$4.2(0.8)$\times$10$^{-5}$ during the
2000 superoutburst and $-$7.3(3.1)$\times$10$^{-5}$ during the 2002
superoutburst.  They agree with each other within the error.

The superhump period of 0.0770 d is near the mode of the $P_{\rm SH}$
distribution (see e.g. \cite{kol99CVperiodminimum, kat03hodel}).  The
delay of the superhump emergence was constrained to be within 2 days
during the 2000 July superoutburst.  This short delay is in accordance
with the relatively long superhump period (see e.g. table 1 in
\cite{osa96review}).  All the values of the amplitude of the
superoutburst, these outburst cycles, the $P_{\rm SH}$ change rate, the
superhump delay, the decline rates of 0.09--0.13 mag d$^{-1}$ during the
plateau phase and of 1.2 mag d$^{-1}$ during the rapid decline phase
are typical values for an SU UMa star having $P_{\rm SH}=0.07700$ d (see
\citep{nog97sxlmi, kat03v877arakktelpucma, war95suuma}.  Note that we
did not detect a significant change in the decline rate throughout the
plateau phase, in contrast to that this rate is expected to become
smaller with depletion of the gas in the outer disk \citep{can01wzsge}.

\begin{table}
\caption{Previous outbursts.}\label{tab:outburst}
\begin{center}
\begin{tabular}{llrlrll}
\hline\hline
\multicolumn{3}{c}{Date$^*$} & $V_{\rm max}$ & D$^\dagger$ & Type$^\ddagger$ & Comment \\
\hline
1994 & Sep. & 25 & 12.8p &       &    & Single Obs. \\
1994 & Dec. & 29 & 14.8p &       &    & Single Obs. \\
1998 & Apr. & 02 & 12.8p &       &    & Single Obs. \\
1999 & Oct. & 04 & 12.2  & $>$11 & S \\
2000 & May  & 05 & 14.9  & 2     & N \\
2000 & Jul. & 05 & 12.4  & 15    & S \\
2001 & Jan. & 25 & 13.2  & 1?    & N \\
2001 & Feb. & 08 & 13.5  & $>$5  & S \\ 
2001 & Apr. & 29 & 13.2  & 2?    & N \\
2001 & Jun. & 18 & 13.7  & 2?    & N \\
2001 & Aug. & 10 & 13.1  & 1?    & N \\
2002 & Mar. & 21 & 13.1  & 2     & N \\
2002 & May  & 16 & 13.9: &       & N? & Single Obs. \\
2002 & May  & 29 & 12.7  & 15    & S  \\
2002 & Aug. & 01 & 13.2  & 2     & N? \\
2002 & Aug. & 08 & 13.4  &       &    & Single Obs. \\
2003 & Feb. & 25 & 12.7  & $>$10 & S \\
2003 & Mar. & 15 & 13.1  & $<$3    & N  & Single Obs. \\ 
2003 & Jun. & 22 & 13.1  & 3     & N \\
\hline
\multicolumn{7}{l}{$^*$ The discovery date.}\\
\multicolumn{7}{l}{$^\dagger$ Duration of the outburst in a unit of day.}\\
\multicolumn{7}{l}{$^\ddagger$ N: normal outburst, S: superoutburst.}
\end{tabular}
\end{center}
\end{table}

We can here estimate the distance to QW Ser by applying the relation
between the orbital period ($P_{\rm orb}$) and the absolute maximum
brightness proposed by \citet{war87CVabsmag}.  The superhump period is
used instead of $P_{\rm orb}$, since the orbital period of QW Ser has
not yet been measured and the superhump period is known to be only a few
percent longer than $P_{\rm orb}$.  The error introduced by this is much
smaller than other factors.  Since the lack of eclipses in the light
curve means that the inclination is not so high, the inclination effect
to the observed flux \citep{war86NLabsmag} should be negligible.
The absolute maximum magnitude is thus expected to $M_V = 5.2 \pm 0.2$
from the Warner's relation.  Then the distance is estimated to be 380
($\pm$ 60) pc, taking into account that this maximum magnitude should be
compared to the apparent maximum magnitude of the normal outburst in the
case of SU UMa-type dwarf novae (cf. \cite{kat02v592her};
\cite{can98DNabsmag}).  This distance is smaller than the secure upper
limit estimated using the proper motion of QW Ser and the maximum
expected velocity dispersion of CVs (\cite{har00DNdistance}).

The X-ray luminosity in the range of 0.5-2.5 keV can be guessed to be
$\log L_{\rm X} = 31.0 \pm 0.1$ from the ROSAT data and the distance,
making use of the formulation given by \citet{ver97ROSAT}.  This
luminosity is a little higher than, but not far from the average value
of SU UMa stars, which is consistent with that QW Ser has typical
properties for an SU UMa-type dwarf nova in other points.

\vskip 3mm

The authors are very thankful to amateur observers for continuous
reporting their valuable observations to VSNET.  Thanks are also to the
anonymous referee for useful comments.  We used the data obtained by the
All Sky Automated Survey project which are kindly opened into public.
This work is partly supported by a Research Fellowship of the Japan
Society for the Promotion of Science for Young Scientists (MU and RI),
and a grant-in-aid from the Japanese Ministry of Education, Culture,
Sports, Science and Technology (No. 13640239, 15037205).

\end{document}